\documentclass[aps,prc,reprint,eqsecnum,showpacs,floatfix]{revtex4-2}
\pdfoutput=1 

\usepackage{amsmath,mathtools}
\usepackage{amssymb}

\usepackage[export]{adjustbox}

\usepackage{xcolor}
\usepackage{graphicx}

\newcommand{\beq}{\begin{equation}} 
\newcommand{\eeq}{\end{equation}} 
  
\newcommand{\bev}{\begin{verbatim}} 
\newcommand{\eev}{\end{verbatim}} 

\newcommand{\vbr}{\boldsymbol{r}}
\newcommand{\vvbbrr}{\boldsymbol{r},\boldsymbol{r'}} 
\newcommand{\ecmcor}{E_{\mathrm{cm}}}
\newcommand{\erotcor}{E_{\mathrm{ rot}}}

\begin{document}

\title{Proton halo structures and $^{22}$Al}

\author{Panagiota Papakonstantinou$^{+1}$, Myeonghwan Mun$^{*1}$, Cong Pan$^\dagger$, Kaiyuan Zhang${^\dagger}{}^\dagger$}

\affiliation{$^+$Institute for Rare Isotope Science, 
                        Institute for Basic Science, Daejeon 34000, Republic of Korea; ppapakon@ibs.re.kr}
\affiliation{$^*$Department of Physics, Kyungpook National University, Daegu 41566, Republic of Korea \\ and Department of Physics and Origin of Matter and Evolution of Galaxies (OMEG) Institute,
Soongsil University, Seoul 06978, Republic of Korea; aa3101@gmail.com} 
\affiliation{$^\dagger$ Department of Physics, Anhui Normal University, Wuhu 241000, China; cpan@ahnu.edu.cn} 
\affiliation{${^\dagger}{}^\dagger$ Institute of Nuclear Physics and Chemistry, China Academy of Engineering Physics, Mianyang, Sichuan 621900, China; zhangky@caep.cn} 

\begin{abstract}

Inspired by the recent debate as to whether the proton drip-line nucleus $^{22}$Al demonstrates a halo structure in its ground state and in order to assess such a possibility, 
we have analyzed theoretical results obtained within the relativistic density functional theory in $^{22}$Al and in a number of neighboring nuclei especially along isotopic, isotonic, and isobaric chains. 
The theory includes self-consistently the effects of pairing, deformation and the continuum. We employ two different functional parameterizations, PC-F1 and PC-PK1.  
Although the valence proton of the $^{22}$Al nucleus is found very loosely bound, in concordance with experimental data, its spatial distribution is found to hardly penetrate the potential barrier. 
Its wave function is found to consist predominately of $\ell =2$ components, for which halo formation is disfavored. 
Comparisons with results for isobars reveal a somewhat more extended density distribution than that of the stable or neutron-rich counterparts, but comparisons along isotopic, isotonic, and isobaric chains reveal no discontinuities in size evolution, which, if present, might have signaled exotic structures. 
\end{abstract}
\maketitle

\maketitle 

\section{Introduction} 

Finite quantum systems, such as nuclei, exhibit special features which are not found in other many-body systems, most promenent among them being shell structure. 
In addition, nuclei are self-bound and governed by the complex nuclear interaction in the non-perturbative regime. 
Its interplay with electroweak interactions determines nuclear stability and the limits of the nuclear landscape. 

At the limits of stability, exotic phenomena may be observed, such as new magic numbers and halo structures~\cite{TSK2013}, whereby one or a few very loosely bound nucleons 
orbit the rest of the system at very large distances on average, {\em i.e.}, well beyond what is expected from systematics. 
One of the best known examples is that of $^{11}$Li~\cite{Tan1985}, assumed to consist of a $^9$Li core surrounded by a two-neutron halo, while $^8$B might be a proton halo nucleus~\cite{War1995}.
A total of about twenty nuclei have been identified as halo nuclei or as candidates for halo structures, see \cite{TSK2013,Zha2023} and references therein. 
The existence and properties of these exquisite quantum structures enrich our understanding of nuclear forces and the physics of loosely bound quantum systems. 


Recently, $^{22}$Al was claimed to be a halo nucleus based on the large asymmetry between the transition from $^{22}$Si to the $1^+$ state of $^{22}$Al on one hand 
and the mirror transition from $^{22}$O to the $1^+$ state of $^{22}$F on the other~\cite{Lee2020}. 
It was pointed out in Ref.~\cite{ZPW2024} that a) the large asymmetry does not necessarily imply a halo structure in the ground state  of $^{22}$Al because it corresponds to 
 excited $1^+$ states and in addition b) the typical experimental signatures for a halo structure, namely enhanced reaction cross sections or narrow momentum distributions~\cite{War1995,Kel1996,War1998,Kan2003}, are not observed. 
It was argued that the large asymmetry could also be attributed to the differences between the parent nuclei, the double magic $^{22}$O and the likely unbound $^{22}$Si. 
The theoretical analysis in Ref.~\cite{ZPW2024} furthermore showed that 
%
the typical condition for halo development, namely a large $s-$component for the weakly bound valence proton, is unlikely to be fulfilled. 
More recently, an analysis of mirror energy differences also does not support the existence of a proton halo in the ground state of $^{22}$Al~\cite{Yu2024}.

Motivated by the above debate as to whether $^{22}$Al is a proton halo nucleus and by the continuing investigation of the proton drip line in experimental facilities, we inspect theoretical structure results for $^{22}$Al in some detail. We employ the deformed relativistic Hartree-Bogoliubov theory in continuum (DRHBc)~\cite{Zhou2010,Li2012,Zhang2020,Pan2022}, 
which self-consistently takes into account pairing, axial deformation, and the continuum and has been employed in several studies of exotic effects including nuclear halos~\cite{ZZM2023,Zhang2023,2023-CNPCa,2023-CNPCb,XLY2023,AnZ2024,ZPW2024,PZZ2024} and also for building a comprehensive nuclear mass table~\cite{Zha2022,Guo2024}. 
The point-coupling PC-PK1 density functional~\cite{Zhao2010} will be used, as in the mass table calculations of \cite{Zha2022,Guo2024} and several other recent sudies of loosely bound nuclei. 
For comparison, we will show selected results also with the PC-F1 density functional~\cite{BMM2002}, which was used in the analysis of Ref.~\cite{ZPW2024}.
We will present selected results also using the triaxial relativistic Hartree-Bogoliubov theory in continuum (TRHBc)~\cite{Zha2023}, to investigate the effects of triaxial deformation. 
We will compare results for $^{22}$Al with those of neigboring nuclei and isotonic and isobaric counterparts in an effort to identify relevant structural changes, if any.

This paper is structured as follows. In Sec.~\ref{sec:halo}, we recall the conditions for halo formation based on elementary quantum-mechanical arguments. 
In Sec.~\ref{sec:drhbc}, we present the theoretical framework used in this work.  
In Sec.~\ref{sec:results}, we analyse results for $^{22}$Al, neighboring nuclei, and isotopic, isotonic, and isobaric counterparts. 
We summarize the conclusions in Sec.~\ref{sec:summary}.  

\section{Recognizing a halo \label{sec:halo}} 

In most nuclei, where the number of neutrons is larger than the number of protons by at least a few units, the neutron distribution is more extended than the proton distribution. 
The difference between the root mean square (r.m.s.) radii of the two distributions defines the neutron skin thickness. 
In proton-rich nuclei and, owing to Coulomb repulsion, in isospin-symmetric nuclei as well, the proton distribution is more extended and one can define analogously the proton skin thickness. 
The skin thickness is expected to vary monotonically along isotopic or isotonic chains and mostly smoothly, except for possible kinks at shell closures--see, e.g., \cite{Guo2024,Mun2024}.   

The halo, whereby the nuclear density extends to far larger distances than what would be expected from mass number systematics, 
is a more extreme phenomenon and is found only in very loosely bound nuclei. 
The basic quantum-mechanical origin of an extended density distribution finds a simple illustration in mean-field theories. 
At moderate to large distances from the nuclear center---beyond, {\em e.g.}, the r.m.s. radius---the wave function of a nucleon with mass $m$ in the mean field potential $U(\vbr )$ decays exponentially as 
$\exp\{-\sqrt{2mE/\hbar^2}r\}$, where $E$ is the nucleon's binding energy. 
Thus the wave function of a loosely bound neutron, with $E$ typically less than $2$~MeV, will decay very slowly compared with more bound nucleons. 

The above expression assumes the vanishing of the mean field at moderate distances from the nuclear center, which is possible for neutrons especially in low-angular momentum states, 
in which case the centrifugal barrier entering the radial Schr\"odinger equation is weak. However, the situation is different for protons:  The Coulomb potential extends to large distances, dropping in value very slowly, so that 
the relevant energy scale for a proton of binding energy $E$ is higher than $E$ by some amount $E_c$ comparable to the Coulomb barrier. 
Then the exponential $\exp\{-\sqrt{2m(E+E_c)/\hbar^2}r\}$ may decay too fast to allow for a proton halo even at almost vanishing energies $E$. 

A comprehensive analysis of halo systems that includes the effects of angular momentum 
can be found in Ref.~\cite{RJM1992}.  
There, it is demonstrated analytically that potentials dropping slower than $1/r^2$, as Coulomb does, will keep nucleons confined. 
In numerical calculations, the core charge density, whose r.m.s. radius we denote by $R_c$, was assumed to have a Gaussian form, appropriate for light nuclei, $\rho_{\rm core} \propto e^{-r^2/b^2}$. 
The r.m.s. radius of the Gaussian distribution equals $R_c=\sqrt{3/2}b$. 
In cases of valence proton energies such that $|E| b^2<1$~MeV~fm$^2$, 
the r.m.s. radius of the valence proton density distribution was found to reach at most a value of roughly $2.7b\approx2.2R_c$ for an $s-$wave, $2b\approx 1.6R_c$ for a $p-$wave, and $1.5b\approx 1.2R_c$ for a $d-$wave. 
For a neutron $s-$wave, on the other hand, the radius can be several times the value of $b$; it can be about $3b$ for a $p-$wave; and $2b$ for a $d-$wave. 
The cleanest halo cases are therefore  possible for neutron $s-$orbitals, which likely is the case in light nuclei in the neutron $p-sd$ region~\cite{TSK2013}. 

Generally, it is not straightforward to characterize a halo based on structure calculations. 
For example, the case of an orbital extending well beyond the core density distribution can also be classified as a halo if it goes beyond systematics, but one has to quantify this argument. 
With the above numerical results as a gauge, it is reasonable to classify any valence nucleon with a distribution mean square radius exceeding one and a half times the core r.m.s. radius as a halo nucleon. 
Although the analysis in \cite{RJM1992} was based on spherical symmetry, the above criterion seems sensible also in the presence of deformation, with some modification. 
Instead of using the total density distribution's moments and the contributions to them from the orbital of interest (in which case only $s-$components contribute)
one can inspect the density distributions along a specific  axis or direction, where the valence orbital is most extended, {\em i.e.} compare direction-specific mean square radii  
\begin{equation}
   r_{\mathrm{rms}}(\hat{r}) = \sqrt{\frac{\int \rho (\mathbf{r}) r^{4} dr }{\int \rho (\mathbf{r}) r^{2} dr}}
\label{eq:dsrms}
\end{equation}  
within an elementary solid angle $d\Omega$ around a chosen direction $\hat{r}$, where $\mathbf{r}=r\hat{r}$.


\section{Theoretical framework\label{sec:drhbc}}

The starting point of the relativistic density functional theory is the effective Lagrangian density with a point-coupling interaction~\cite{Meng2016}, 
\begin{eqnarray}
{\cal L} &=& \bar{\psi} \left(i\gamma_\mu \partial^\mu - m \right)\psi\,
-\frac{1}{2} \, \alpha_{S} \left(\bar{\psi}\psi \right) \left(\bar{\psi}\psi \right)\,
  \nonumber \\ & & 
-\frac{1}{2} \, \alpha_{V} \left(\bar{\psi} \gamma_\mu \psi \right) \left(\bar{\psi} \gamma^\mu \psi \right)\,
- \frac{1}{2} \, \alpha_{TV} \left(\bar{\psi} \vec{\tau} \gamma_\mu \psi \right)
 \left(\bar{\psi} \vec{\tau} \gamma^\mu \psi \right)\,                         
  \nonumber \\ & & 
-\frac{1}{2}\alpha_{TS} \left(\bar{\psi} \vec{\tau} \psi \right) \left(\bar{\psi} \vec{\tau} \psi \right)
 \nonumber \\   & &                           
-\frac{1}{3}\beta_S \left(\bar{\psi}\psi \right)^3
-\frac{1}{4}\gamma_S \left(\bar{\psi}\psi \right)^4
-\frac{1}{4} \, \gamma_{V} \left[\left(\bar{\psi} \gamma_\mu \psi \right)
\left(\bar{\psi} \gamma^\mu \psi \right)\right]^2\,   
\nonumber \\  & & 
-\frac{1}{2} \delta_S \partial_\nu \left(\bar{\psi}\psi \right)
 \partial^\nu \left(\bar{\psi}\psi \right)
-\frac{1}{2} \delta_V \partial_\nu \left(\bar{\psi} \gamma_\mu \psi \right)
 \partial^\nu \left(\bar{\psi} \gamma^\mu \psi \right)
  \nonumber \\    & & 
-\frac{1}{2} \delta_{TV} \partial_\nu \left(\bar{\psi} \vec{\tau} \gamma_\mu \psi \right)
 \partial^\nu \left(\bar{\psi} \vec{\tau} \gamma_\mu \psi \right)                                                       
  \nonumber \\ & & 
-\frac{1}{2}\delta_{TS} \partial_\nu \left(\bar{\psi} \vec{\tau} \psi \right) \partial^\nu\left(\bar{\psi} \vec{\tau} \psi \right)
  \nonumber \\ & & 
-\frac{1}{4} F^{\mu \nu} F_{\mu \nu} - e  \bar{\psi} \gamma^\mu \frac{1-\tau_3}{2} A_\mu \psi  ,
\label{Lag}
\end{eqnarray}
where $m$ denotes the nucleon mass and 
$\alpha_{S}$, $\alpha_{V}$, $\alpha_{TV}$, and $\alpha_{TS}$ denote the coupling constants for four-fermion contact interactions. The terms involving  $\beta_{S}$, $\gamma_{S}$ and $\gamma_{V}$ account for medium effects, those with  $\delta_{S}$, $\delta_{V}$, $\delta_{TV}$, and $\delta_{TS}$   reflect finite-range effects, and $A_\mu$ and $F_{\mu \nu}$  correspond to the four-vector potential and the electromagnetic field strength tensor, respectively. The subscripts $S, \, V$ and $T$  stand for scalar, vector, and isovector, respectively.


By performing the Legendre transformation on the Lagrangian density in Eq.~(\ref{Lag}) and applying the mean-field approximation, we can derive the energy density functional. 
By treating the mean fields and pairing correlations self-consistently, we then arrive at the relativistic Hartree-Bogoliubov equation 
\begin{equation} 
\left(
\begin{array}{cc}
h_D - \lambda & \Delta \\ - \Delta^* & -h^*_D + \lambda
\end{array}
\right) 
\left(
\begin{array}{c}
U_k \\ V_k
\end{array}
\right)
= E_k \,
\left(
\begin{array}{c}
U_k \\ V_k
\end{array}
\right)
\label{eq:hb}
\end{equation}
Here, $E_k$ denotes the quasiparticle energy,
$U_k$ and $V_k$ are the quasiparticle wave functions, with $\lambda$ denoting the Fermi energy.
The Dirac Hamiltonian $h_D$ is given by
\begin{equation}
h_D (\boldsymbol{r}) = \boldsymbol{\alpha} \cdot \boldsymbol{p} \, + \, \beta \left(M+S(\vbr )\right) \, + \, V(\vbr ) ,
\end{equation}
where the scalar $S(\vbr )$ and vector $V(\vbr )$ potentials  can be expressed as
\begin{eqnarray}
S(\vbr ) &=& \alpha_S \rho_S \,+\, \beta_S \rho^2_S \,+\,
          \gamma_S \rho^3_S \,+\, \delta_S \Delta\rho_S, \label{sPot} \\
V(\vbr ) &=& \alpha_V \rho_V \,+\, \gamma_V \rho^3_V \,+\, \delta_V \Delta \rho_V
          \,+\, e A_0 \,   \nonumber  \\   && +\, \alpha_{TV} \tau_3 \rho_{TV} \,+\, \delta_{TV} \tau_3 \Delta \rho_{TV}  . \label{vPot}
\end{eqnarray}
The local densities $\rho_S(\vbr )$, $\rho_V(\vbr )$ and $\rho_{TV}(\vbr )$
 can be expressed in terms of the quasiparticle wave functions as follows:
\begin{eqnarray}
\rho_S(\vbr ) = \sum_{k>0} \, \bar{V_k}(\vbr ) V_k(\vbr ), \\
\rho_V(\vbr ) = \sum_{k>0} \, V^\dag_k (\vbr ) V_k(\vbr ), \\
\rho_{TV}(\vbr ) = \sum_{k>0} \, V^\dag_k (\vbr ) \tau_3  V_k(\vbr ),
\end{eqnarray}
where $k>0$ corresponds to the no-sea aproximation. 
The pairing potential $\Delta$ is given by 
\begin{eqnarray}
\Delta (\vvbbrr ) = V^{pp} (\vvbbrr ) \kappa (\vvbbrr ) , 
\end{eqnarray}
where the pairing tensor is defined by $\kappa=V^{\ast}U^T$. 
For the pairing interaction $V^{pp}$  we use the density-dependent zero-range form 
\begin{eqnarray}
V^{pp} (\vvbbrr ) = \frac{V_0}{2} \left(1 - P^{\sigma} \right)
                             \delta (\vbr - \vbr' )
                             \left(1 - \frac{\rho(\vbr )}{\rho_{\mathrm{sat}}} \right),
\end{eqnarray}
where $V_0$ is the strength parameter of the pairing interaction and $\rho_{\mathrm{sat}}$ is the saturation density of symmetric nuclear matter.

To address axial deformation while maintaining spatial reflection symmetry, in DRHBc theory, the  potentials ($S(\vbr )$, $V(\vbr )$) and densities ($\rho_S(\vbr )$, $\rho_V(\vbr )$, $\rho_{TV}(\vbr )$) are expanded  in terms of Legendre polynomials.
For more details on DRHBc, we refer to~\cite{Li2012,Zhang2020, Pan2022}. 
In the case of TRHBc, the potentials and densities are expanded in terms of spherical harmonics of even angular momentum and magnetic quantum numbers, such that triaxial deformation degrees of freedom are included while spatial reflection symmetry and simplex symmetry are imposed~\cite{ZZM2023}. 
To investigate exotic nuclear properties, it is essential to self-consistently incorporate both continuum and deformation effects besides pairing. 
To account for continuum effects, the eigenstates of the Dirac Woods-Saxon potential placed inside a large box~\cite{ZMR2003,Zhang2022}, are used as a basis to solve the relativistic Hartree-Bogoliubov equations (\ref{eq:hb}). 

The total energy of a nucleus can be expressed as
\begin{eqnarray}
E_{\rm tot}&=& \sum_{k>0}(\lambda_{\tau}\! - \! E_k)v_k^2 - E_{\mathrm{pair}} 
\nonumber \\ & & 
-\int \mathrm{d}^3\vbr \left[ \frac{1}{2}\alpha_S\rho^2_S \right.   +
\frac{1}{2}\alpha_V\rho_V^2 + \frac{1}{2}\alpha_{TV}\rho_{TV}^2 
\nonumber \\ & & 
+ \frac{2}{3}\beta_S\rho_S^3 
+\frac{3}{4}\gamma_S\rho_S^4 +\frac{3}{4}\gamma_V\rho_V^4
\nonumber \\ & & 
  + \! \frac{1}{2} ( \delta_S\rho_S\Delta\rho_S \!
+ \! \delta_V\rho_V\Delta\rho_V\! 
\nonumber \\ & & 
+\! \left. {\textcolor{white}{\frac{a}{b}}} \!\!\!\!\!\delta_{TV}\rho_{TV}\Delta\rho_{TV}\! +\! \rho_peA^0) \right]    +\ecmcor\, ,
\end{eqnarray}
where $\ecmcor$ denotes the center-of-mass correction energy.
The zero-range pairing force results in a local pairing field $\Delta(\mathbf{r})$ with the associated pairing energy energy expressed as follows:
\begin{equation}
E_{\mathrm{pair}} = -\frac{1}{2}\int\mathrm{d}^3\vbr \, \kappa(\vbr )\Delta(\vbr ).
\end{equation}
Finally, a rotational correction energy $\erotcor$ is also considered. 

\section{Results and discussion\label{sec:results}} 

The properties of $^{22}$Al are calculated using the point-coupling PC-PK1 density functional~\cite{Zhao2010}, which is also used in the mass table calculations \cite{Guo2024} and several other recent sudies of loosely bound nuclei, and the PC-F1  density functional~\cite{BMM2002}, which was used in the analysis of Ref.~\cite{ZPW2024}.
The ground state of $^{22}$Al within DRHBc is predicted prolately deformed with both the PC-PK1 and PC-F1 density functionals, with quadrupole deformation parameter $\beta_2=0.29$ and 
$0.31$, respectively.  
Within TRHBc, the result persists in the case of PC-PK1, while in the case of the PC-F1 functional, weak triaxiality is predicted, with $\beta_{20}=0.31$ and $\beta_{22}=0.03$.  
The ground state energies, without taking into account center-of-mass and rotational corrections, are obtained from PC-PK1 within DRHBc, from  PC-F1 within DRHBc and from PC-F1 within TRHBc as $-140.63$, 
$-140.41$~MeV,  and $-140.42$~MeV, respectively. 

Additional results for $^{22}$Al and some neighboring nuclei are tabulated in Table \ref{tab:energies} for reference. 
In particular, we include results with the PK1 functional~\cite{LMG2004}, which is not of the point-coupling type and contains non-linear terms, for a more comprehensive picture of the sensitivity of results with respect both to the spatial symmetry and to the functional form. 
%
\begin{table}
\caption{Energies of the indicated nuclides as predicted by the DRHBc theory using the point-coupling functional PC-PK1 and by the TRHBc theory using PK1~\cite{LMG2004}, and compared with the Atomic Mass Evaluation 2020 (AME2020) \cite{AME2020}, in units of MeV. 
Also shown are the quadrupole deformation parameters $\beta_2$ in DRHBc and $\beta_{20},\beta_{22}$ in TRHBc. 
$E_{\mathrm{tot}}$ is the total kinetic and potential energy excluding the  rotational energy correction ($\erotcor$). 
Results with the TRHBc and PC-PK1 are not shown, because they are practically the same as for DRHBc except for $^{23}$Si.
\label{tab:energies}} 
\begin{center} 
\begin{tabular}{r|cccc} \hline \\[-2mm]
                                                               &      $^{22}_{13}$Al$_9$          &    $^{21}_{12}$Mg$_9$     &   $^{23}_{14}$Si$_9$      &         $^{23}_{13}$Al$_{10}$      \\
\hline\\[-4mm]
   DRHBc, PC-PK1  \hspace{7mm} \mbox{~}
 & & & & \\
  $E_{\mathrm{tot}}\!\!-\!\!\ecmcor$             &        140.63          &     140.17          &        142.60        &      158.13               \\
     $E_{\mathrm{tot}}$                       &         150.01            &     149.29           &        151.42         &         167.58               \\
     $+\erotcor$                         &       152.10          &      151.03          &      155.18    &         170.24                \\
     $\beta_2$                                         &           0.29            &      0.33             &          -0.18        &       0.35 \\
\hline
     AME2020                                           &          149.204        &         149.205     &        150.742      &     168.722 \\ 
   \hline
  TRHBc, PK1  \,  $E_{\mathrm{tot}}\!\!-\!\!\ecmcor $            &     139.39          &     138.93           &        140.87         &        156.93               \\
   $\beta_{20},\beta_{22}$                &  $ 0.28, 0$            &    $0.34,0$    & $0.10,0.12$       &   $0.39,0$    \\
\hline
\end{tabular}
\end{center} 
\end{table}
%
%
These results show minor variations with respect to the method of calculation except for $^{23}$Si whose triaxial deformation is predicted substantial 
by the PK1 density functional.
\begin{figure}[b]
\centerline{
\includegraphics[width=0.43\textwidth]{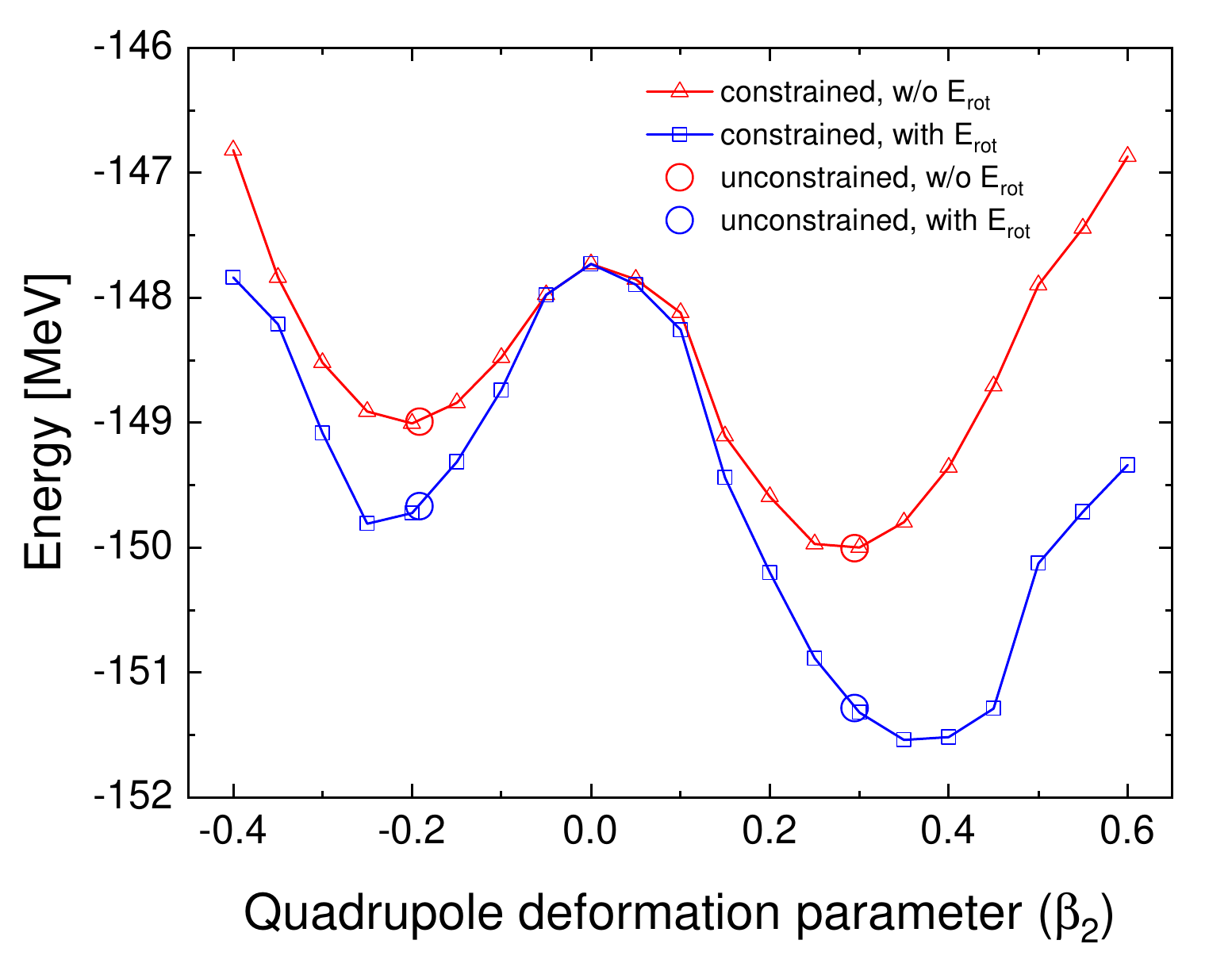} } 
\caption{Potential energy curve for $^{22}$Al, with or without the rotational correction energy (``E$_{\mathrm{rot}}$") included, 
generated by performing constrained DRHBc calculations and using the PC-PK1 energy density functional. 
The results of unconstrained calculations initialized to different variational parameters are also shown.
\label{fig:Al22constr}} 
\end{figure} 

Inspecting the potential energy curve of $^{22}$Al, Fig.~\ref{fig:Al22constr}, which is generated by performing constrained calculations, 
we find that there is also an oblate local minimum close in energy to the prolate minimum.
Although we show the curve for the PC-PK1 functional, the same holds for the PC-F1 functional. 
(Note that the rotational correction energy $\erotcor$ is not considered in the variational calculations.  
When we add $\erotcor$ after variation, the minimum may slightly shift, as observed here.) 
We now proceed to examine the canonical single-particle spectra and the density distributions of both solutions. 

\subsection{Energetics and density distribution of $^{22}$Al } 

\begin{figure} \centerline{
\includegraphics[width=0.25\textwidth,valign=c]{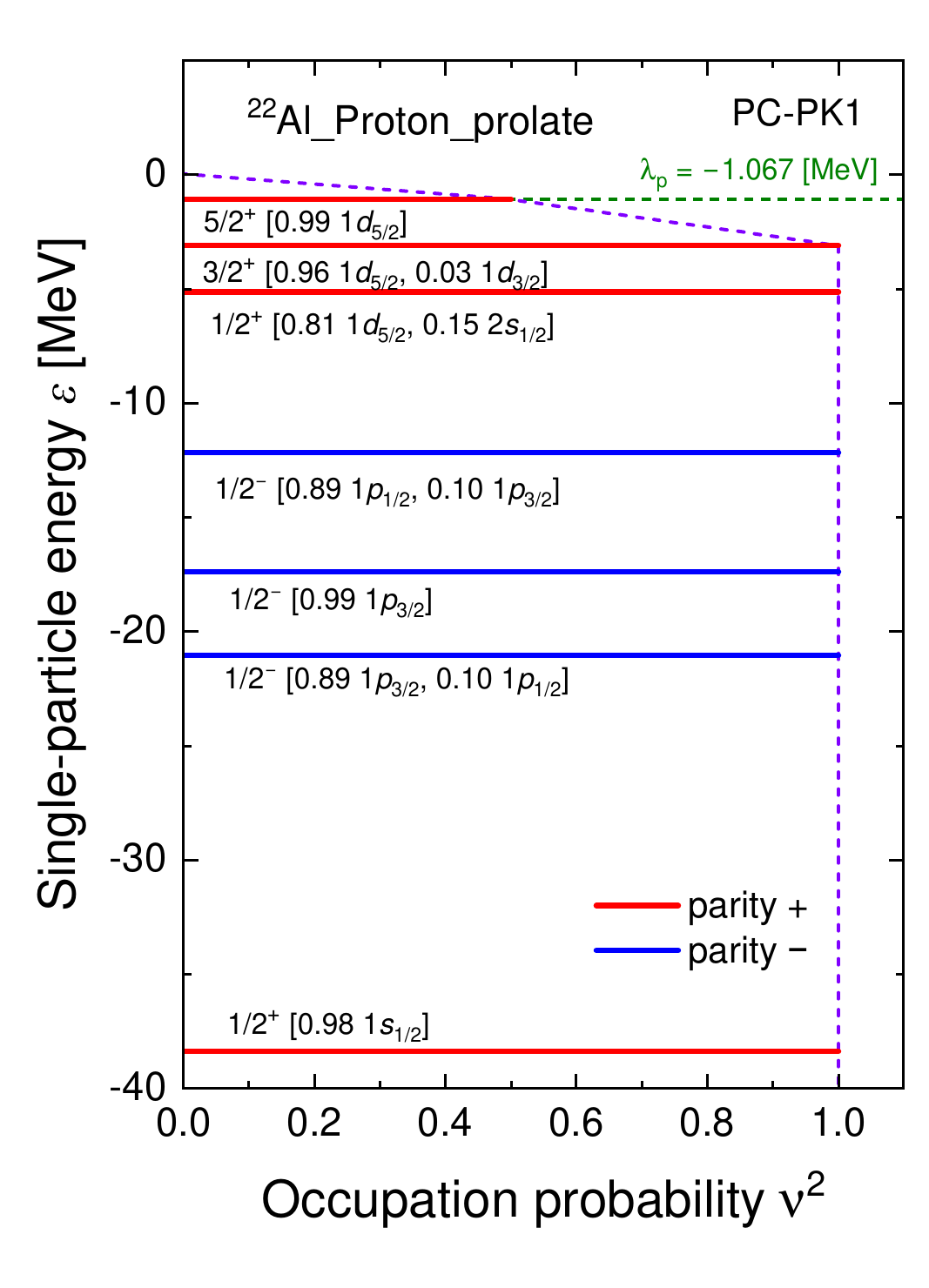}
\includegraphics[width=0.25\textwidth,valign=c]{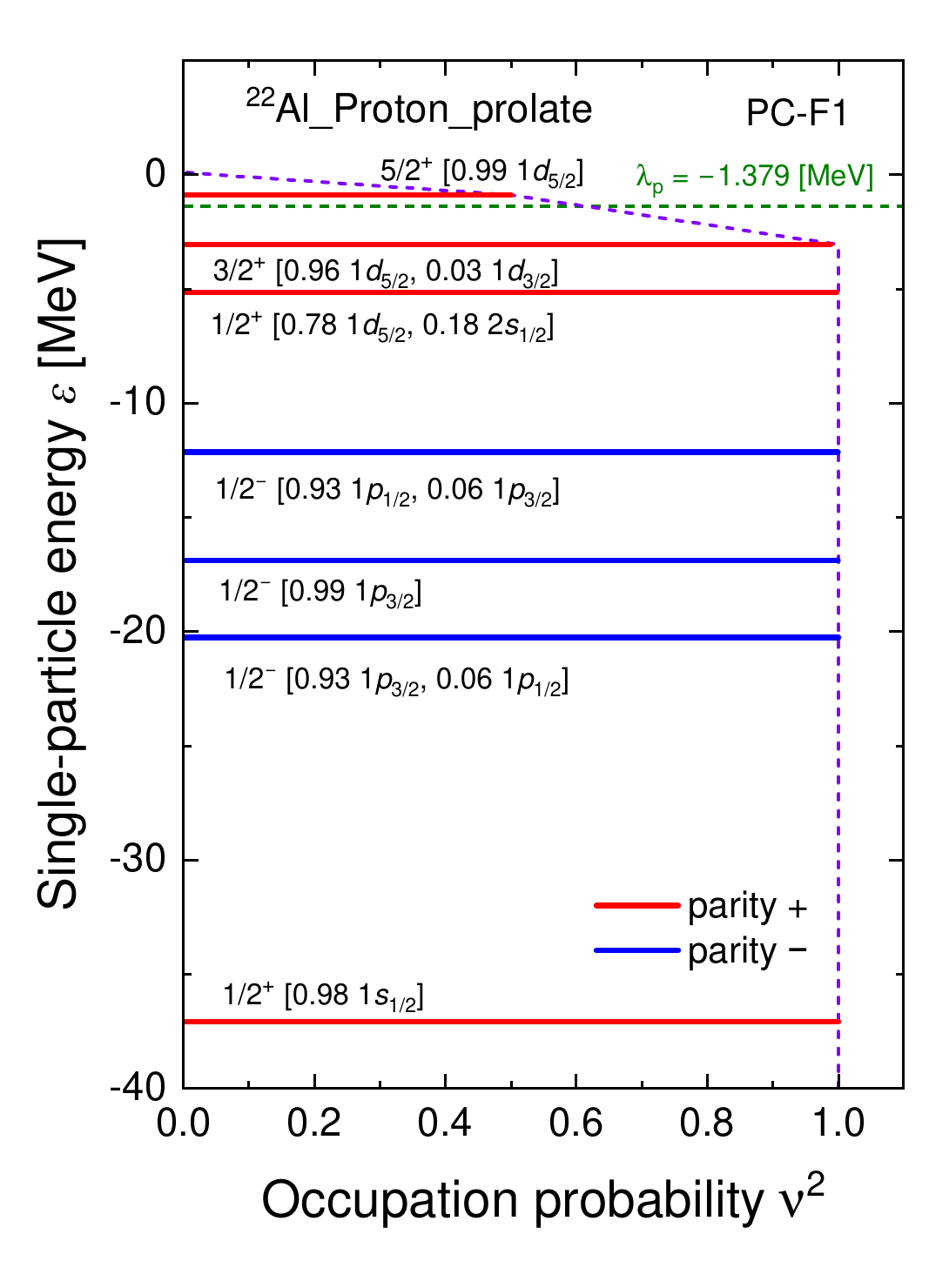}
}
\caption{DRHBc results for the canonical proton quasiparticle energies and occupation probabliities using the PC-PK1 and PC-F1 functionals in the prolate state of $^{22}$Al.
Major $n \ell j$ orbital contributions are indicated. 
A dashed line marks the Fermi energy, whose value is indicated. 
The connecting dotted lines are drawn to guide the eye.
\label{fig:SPEs} } 
\end{figure}

The canonical single-particle energies of the prolate ground state of $^{22}$Al calculated within DRHBc and their occupation probabilities are shown in Fig.~\ref{fig:SPEs} for both PC-PK1 and PC-F1 functionals. 
Also indicated are the major contributions to each $m^{p}$ state of $\ell j$ orbitals. 
The least bound proton state is at about $-1$~MeV. 
All of the less bound proton orbitals are predominately $d$ states, with no $s$ contribution. 

\begin{figure*}
\centerline{
\includegraphics[width=0.90\textwidth]{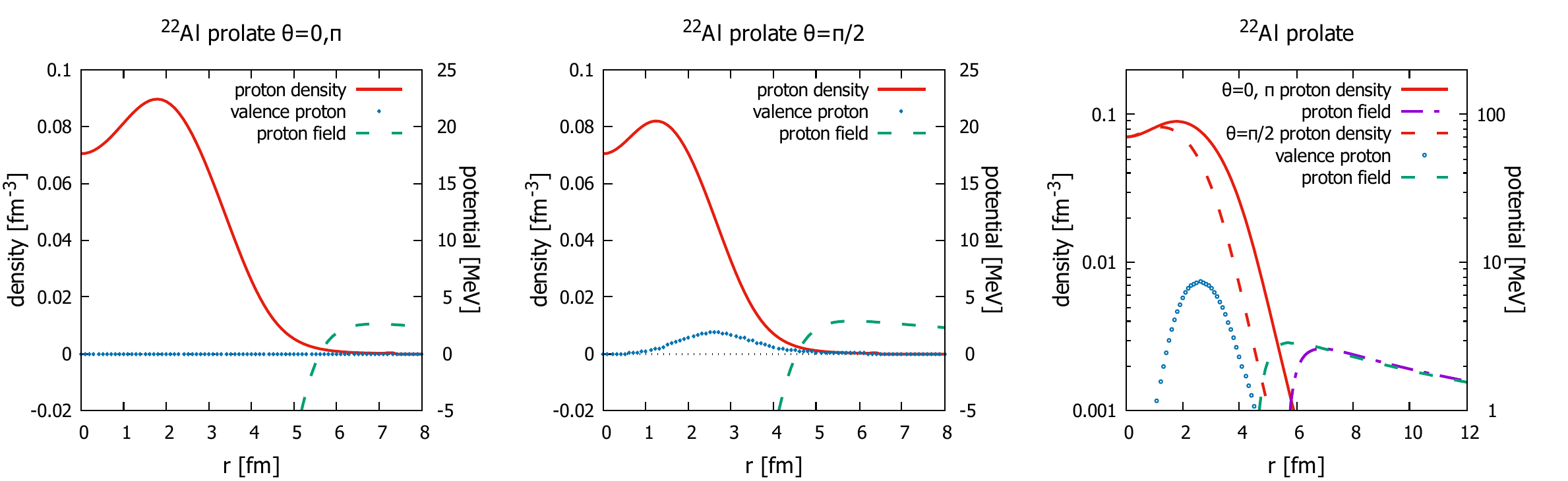} } 
\caption{DRHBc results for the prolate ground state of $^{22}$Al. 
Shown are the proton density distribution, the valence proton contribution, and the proton field along the $z-$axis ($\theta=0,\pi$) and perpendicular to it ($\theta=\pi/2$), 
in linear (left, middle) and logarithmic (right) scale, where the valence proton density along the $z$  axis lies below the displayed range.
\label{fig:profiles}} 
\end{figure*}

The proton field calculated within DRHBc along the $z-$axis and perpendicular to it in the case of $^{22}$Al is shown in Fig.~\ref{fig:profiles} for PC-PK1. 
The positive-valued tail comes from the Coulomb potential. 
The proton density distribution along the $z-$axis and perpendicular to it and the contribution from the valence proton are also shown.
One can see that the Coulomb potential remains well above $1$~MeV for a long distance. 
The valence proton density is situated at the nuclear surface near the $xy-$plane. 
Generally, the tail of the density distribution hardly penetrates the positive part of the mean field. 

As a quantitative way to assess the nature of the valence orbital, we turn to the direction-specific r.m.s. defined in Eq.~(\ref{eq:dsrms}). 
We calculate this quantity along the axis perpendicular to $z$, $\theta=\pi/2$, for the total density distribution, $r_{\rm tot}$, 
the contribution of the valence proton to the density distribution, $r_{\rm val}$, and for the difference of the two distributions, which plays the role of the core distribution, $r_{\rm cor}$. 
We find
\[ 
  r_{\rm tot}=2.88~{\rm fm},   r_{\rm cor}=2.70~{\rm fm},   r_{\rm val}=3.60~{\rm fm}.
\]  
Thus we get $r_{\rm val}=1.33r_{\rm cor}$, consistent with the estimate of Ref.~\cite{RJM1992} for loosely bound protons, as discussed in Sec.~\ref{sec:halo}, and not indicative of a halo structure. 
Note that the total r.m.s. charge radius of the $^{21}$Mg ground state, which is the system with one proton less, is found to be $2.90$~fm, {\em i.e.}, larger than $r_{\rm cor}$.  

\begin{figure*} \centerline{
\includegraphics[width=0.70\textwidth]{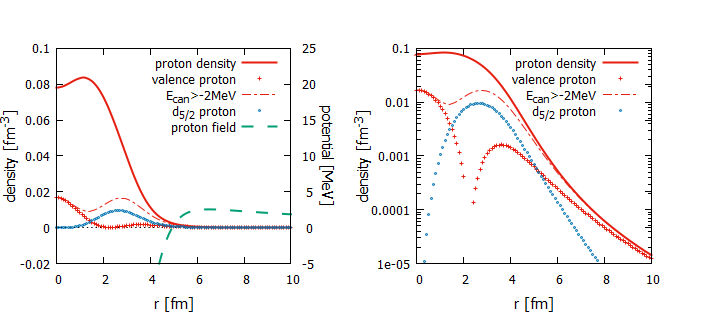}
}
\caption{DRHBc results for the spherical ground state of $^{21}$Al. 
Shown are the proton density distribution, the valence proton contribution ($2s_{1/2}$), 
the contribution of states with canonical energy above $-2$~MeV, separately also the contribution of the $1d_{5/2}$ state, and the proton field. 
The right panel is in logarithmic scale.
\label{fig:Al21} } 
\end{figure*}

One may ask whether proton halo formation requires the valence proton's energy to be positive. 
To illustrate the point, let us inspect the results for the recently observed proton-unbound $^{21}$Al~\cite{Kos2024}, which is predicted spherical. 
It is also predicted correctly to be particle-unbound by our calculation. 
The energy of the valence $2s_{1/2}$ proton is found equal to $0.287$~MeV. The energy of the next least bound state, the $1d_{5/2}$  occupied by $3.76$ protons, equals $-1.211$~MeV, which is higher than the valence proton's in $^{22}$Al. 
As Fig.~\ref{fig:Al21} illustrates, the $1d_{5/2}$ is well localized near the nuclear surface, similar to $^{22}$Al. 
The $2s_{1/2}$ state extends to large radii and results in a tail in the density distribution invoking a halo. 
The difference with $^{22}$Al is clear. 
Even so, the contribution of the tail region to the occupation probability is found to be very small. Its existence might be interpreted as the onset of proton drip.

Given that the highest occupied orbital in $^{21}$Al is technically unbound, the results may be sensitive to the size of the box used in our calculations and not well converged. 
However, the general point we wish to make stands, namely, that the Coulomb barrier prevents the formation of a proton halo in the drip-line isotope.  

\begin{figure} \centerline{
\includegraphics[width=0.25\textwidth,valign=c]{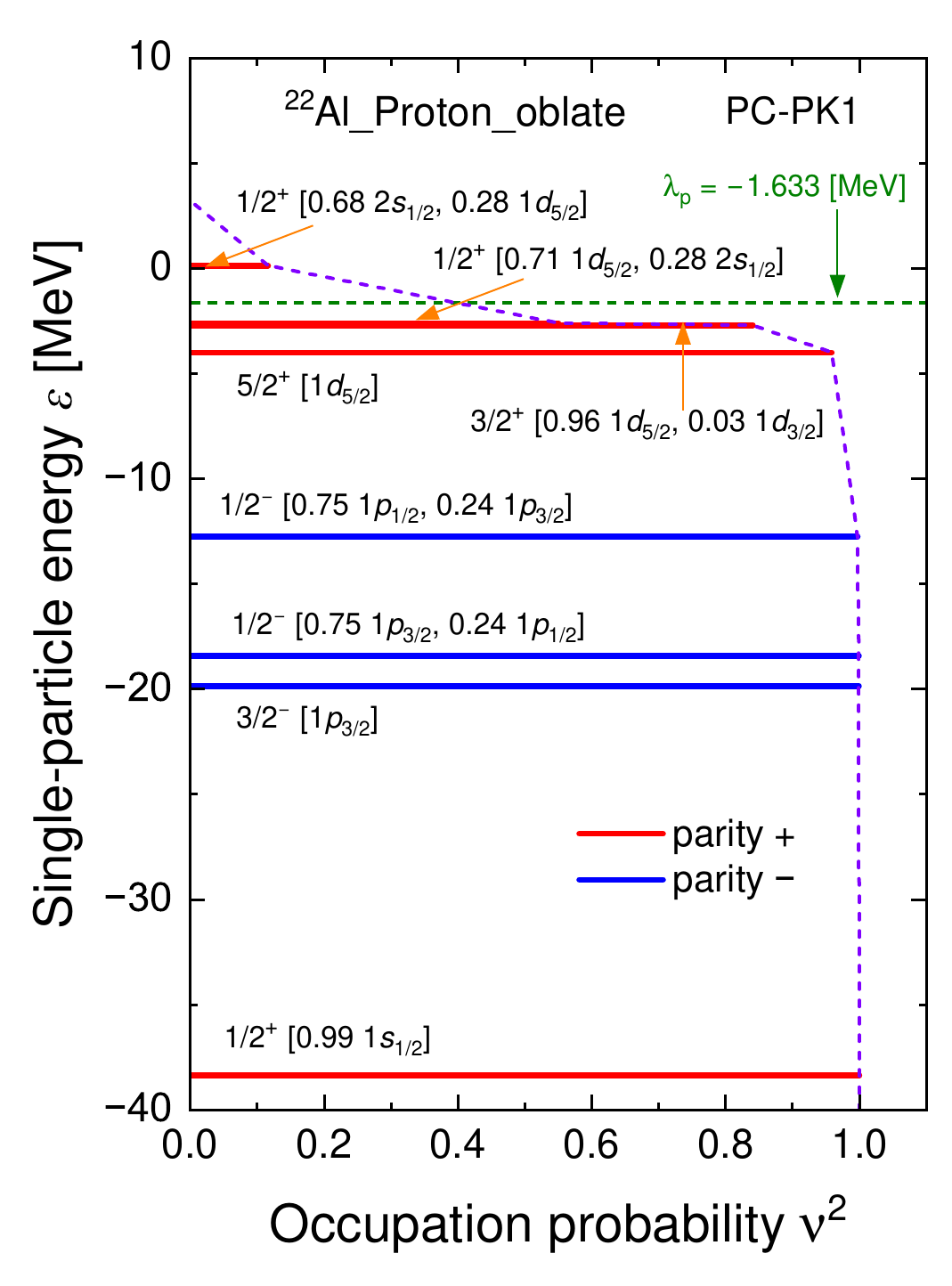}
\includegraphics[width=0.25\textwidth,valign=c]{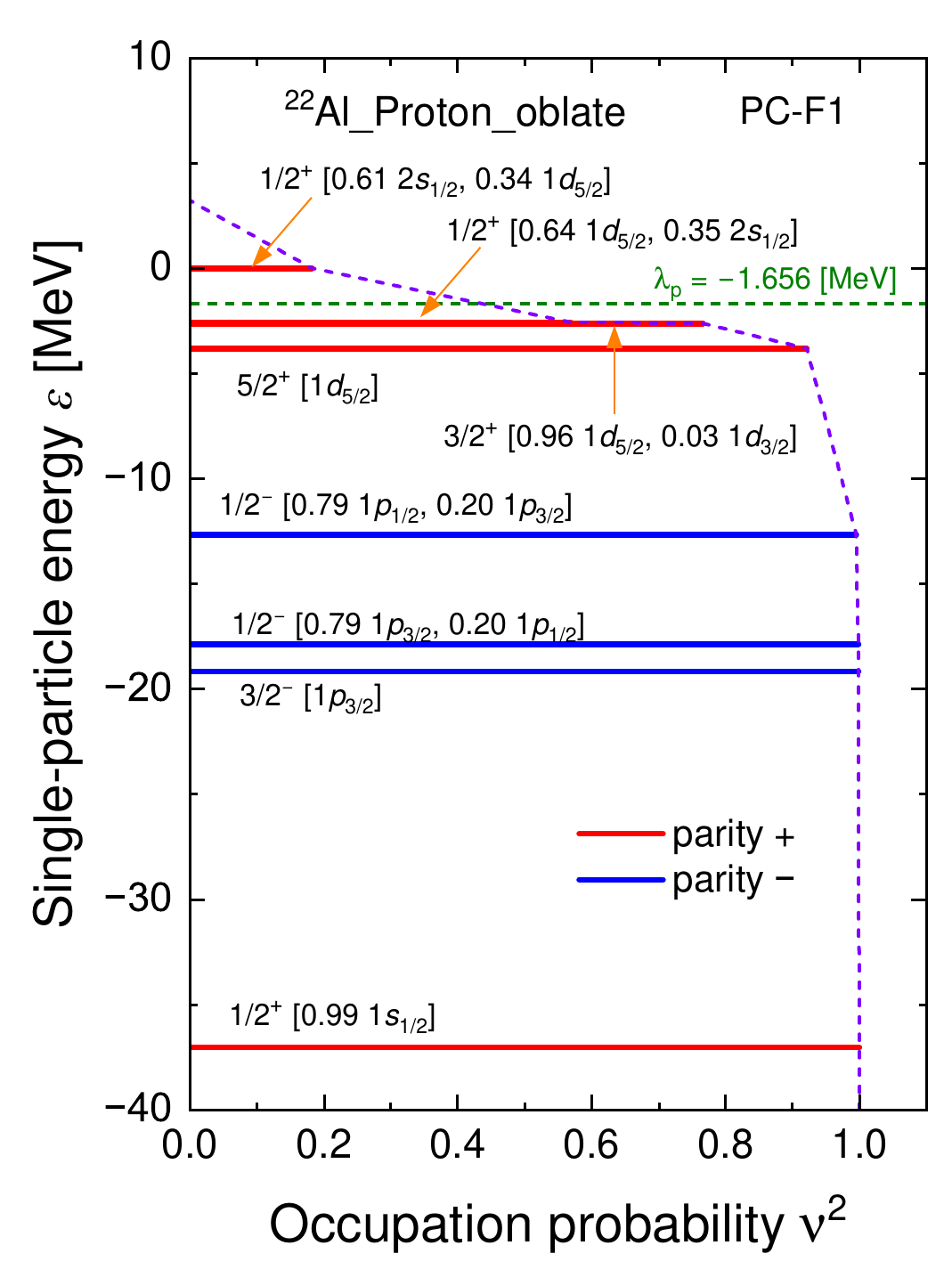}
}
\caption{DRHBc results for the canonical proton quasiparticle energies and occupation probabliities using the PC-PK1 and PC-F1 functionals in the oblate state of $^{22}$Al.
Major $n \ell j$ orbital contributions are indicated. 
A dashed line marks the Fermi energy, whose value is indicated. 
The connecting dotted lines are drawn to guide the eye.
\label{fig:SPEsObl} } 
\end{figure}

Let us finally turn to the oblate solution seen in Fig.~\ref{fig:Al22constr}, 
whose proton separation energy is predicted positive with respect to both the prolate and the oblate states of $^{21}$Mg according to existing calculations with PC-PK1~\cite{Guo2024}. 
The single-particle energies and occupation probabilities of the oblate state of $^{22}$Al are displayed in Fig.~\ref{fig:SPEsObl} for both functionals. 
We observe that there is a proton state with occupancy $v^2=0.12$ ($0.18$) at an energy of $130$~keV (practically zero) for PC-PK1 (PC-F1). 
We may focus on this state since all others lie below $-2$~MeV and using the case of PC-PK1. 
\begin{figure*}
\centerline{
\includegraphics[width=0.90\textwidth]{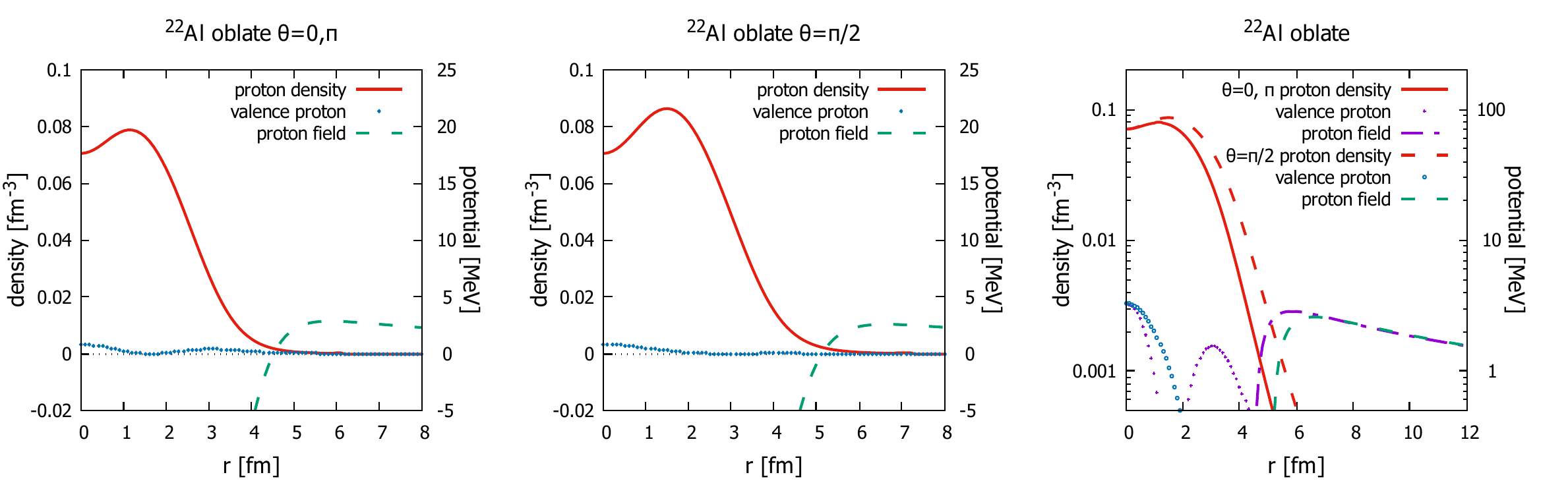} } 
\caption{DRHBc results for the oblate state of $^{22}$Al. 
Shown are the proton density distribution, the valence proton contribution, and the proton field along the $z-$axis ($\theta=0,\pi$) and perpendicular to it ($\theta=\pi/2$), 
in linear (left, middle) and logarithmic (right) scale.
\label{fig:profilesObl}} 
\end{figure*} 
The results for this orbital, as well as the total density distribution and the potential are shown in Fig.~\ref{fig:profilesObl} using the same conventions as for the prolate ground state, Fig.~\ref{fig:profiles}.
We observe that there is hardly any occupancy or tail in the region of the Coulomb barrier. 
For the direction-specific radii we find
\[ 
  r_{\rm tot}=2.78~{\rm fm},   r_{\rm cor}=2.69~{\rm fm},   r_{\rm val}=4.03~{\rm fm}  
\]  
with $r_{\rm val}=1.45r_{\rm cor}$ along the $z-$axis 
and 
\[ 
  r_{\rm tot}=3.20~{\rm fm},   r_{\rm cor}=3.19~{\rm fm},   r_{\rm val}=4.51~{\rm fm} 
\]  
with $r_{\rm val}=1.41r_{\rm cor}$ perpendicular to the $z-$axis.
These values are again consistent with loosely bound protons but not a halo structure. 

Next, we will look at the $^{22}$Al proton and neutron density distributions in comparison with isobars and neighboring nuclei for a more complete picture.

\subsection{A=22 isobars} 

\begin{figure*}
\centerline{
\includegraphics[width=0.70\textwidth]{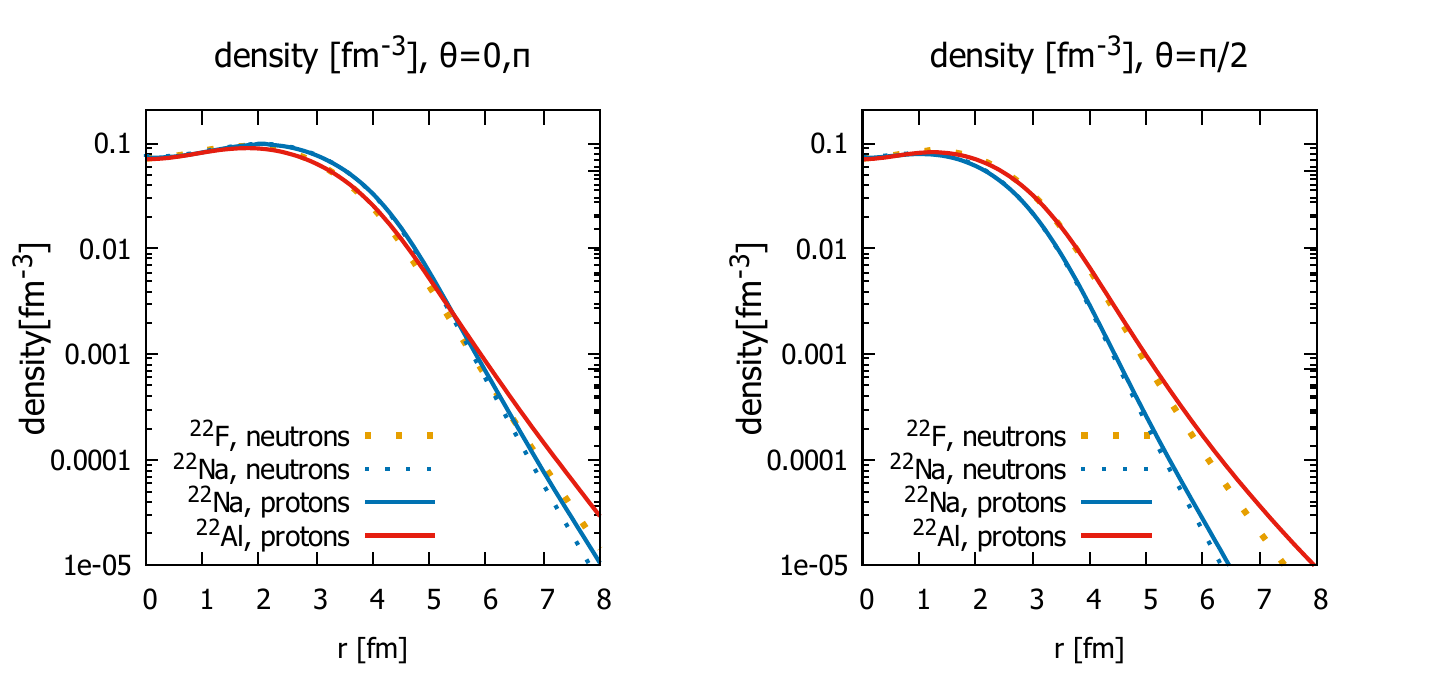} } 
\caption{DRHBc results for the ground-state proton and neutron density distributions of $^{22}$Al, $^{22}$Na, and $^{22}$F along the $z-$axis ($\theta=0$ or $\pi$) and perpendicular to it ($\theta=\pi/2$). 
\label{fig:isobars}} 
\end{figure*}

The mirror nucleus of $^{22}$Al is $^{22}$F with nine protons and thirteen neutrons. 
According to the DRHBc calculations using the PC-PK1 functional, the ground state is also prolate in shape and the neutron and proton Fermi energies equal, respectively, $-6.36$ and $-15.87$ MeV. 
At $-6.32$~MeV, the valence neutron in $^{22}$F is significantly more bound than the valence proton in $^{22}$Al, even if we take into account the extended Coulomb barrier in $^{22}$Al. 
According to the DRHBc results, the densities are almost mirrors of each other, as shown in Fig.~\ref{fig:isobars}, up to a radius of about $5$~fm, where the extended tail of the $^{22}$Al proton distribution shows a clearly different slope. 
The proton and neutron densities of the stable, isospin-symmetric $^{22}$Na are shown for comparison and reference.
They are very close to each other, a small difference arising, as in all isospin-symmetric nuclei, from the protons' Coulomb repulsion.
The $^{22}$F neutron distribution and the $^{22}$Al proton distribution are almost mirrors of each other at small $r$. 
Their tails at large $r$ are more extended than the $^{22}$Na distributions, especially perpendicular to the $z-$axis, but 
the $^{22}$Al proton tail is in turn more extended than the $^{22}$F neutron distribution. 

All the above results are qualitatively consistent with the different binding of the valence nucleons in each nucleus as discussed in Sec.~\ref{sec:halo}. 
In quantitative terms, the calculated proton and neutron r.m.s. radii in $^{22}$Al equal $3.077$ and $2.735$~fm and the neutron and proton radii in $^{22}$F equal $3.001$ and $2.738$~fm, respectively. 
Thus the proton skin thickness in the former equals $0.342$~fm and the neutron skin thickness equals $0.263$~fm in the latter. 
The difference is significant. 
In order to settle whether the extended distribution in $^{22}$Al qualifies as a special structure, we turn to its neighbors on the nuclide chart. 


\subsection{Neighboring nuclides}

\begin{figure*}
\centerline{
\includegraphics[width=0.9\textwidth]{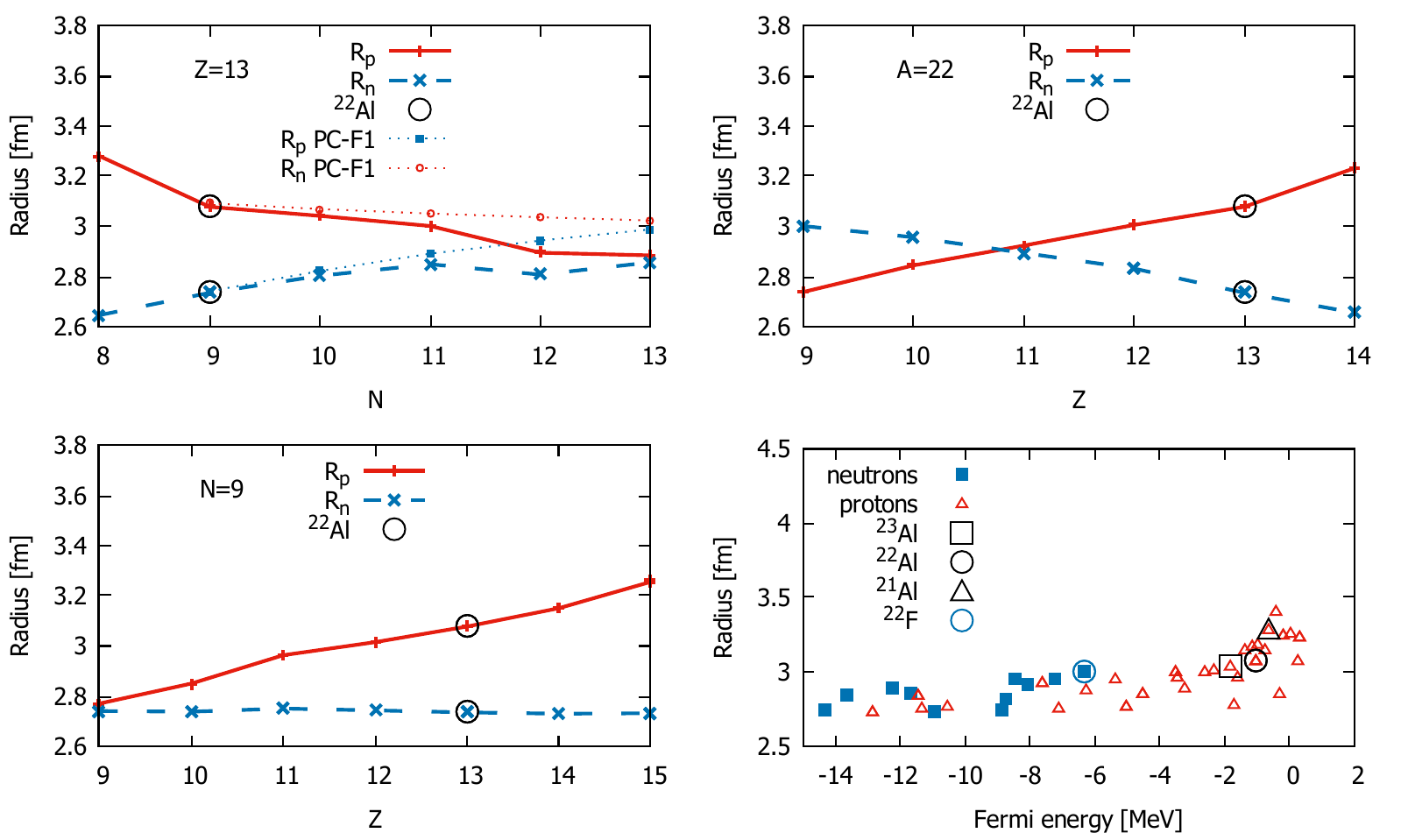} } 
\caption{DRHBc results for the proton and neutron r.m.s. radii 
of $Z=13$ isotopes and $N=9$ isotones in the vicinity of $^{22}$Al as well as $A=22$ isobars using the PC-PK1 functional. 
For $Z=13$ isotopes TRHBc results with the PC-F1 functional are also shown, as indicated.  
The bottom-right panel shows a scatter plot of proton and neutron r.m.s. radii of a number of $Z=9-15$ nuclides in the neighborhood of $^{22}$Al, including some beyond the proton drip line.  
$^{21-23}$Al and $^{22}$F are marked by the indicated large open symbols. 
Results for $^{22}$Al are circled in all panels.  
\label{fig:mytable}} 
\end{figure*}

The experimental data and DRHBc results for the ground state energies of the particle-bound immediate neighbors are given in Table~\ref{tab:energies}. 
In addition to the tabulated nuclides, we have calculated the ground state properties of a number of $Z=9-15$ nuclides in the vicinity of $^{22}$Al, 
some of which, including $^{21}$Al, lie beyond the proton drip line. 
Systematics for the proton and neutron r.m.s. radii are demonstrated in Fig.~\ref{fig:mytable}. 
Radii are plotted along the $Z=9$ isotope, $A=22$ isobar, and $N=9$ isotone chains using the DRHBc results with the PC-PK1 functional. 
For $Z=13$ isotopes TRHBc results with the PC-F1 functional are also shown. 
The results for $^{22}$Al are circled. 
We observe that the trends are smooth as we go from the line of stability to $^{22}$Al, while kinks are observed only beyond $^{22}$Al on the proton-rich side. 
Proton and neutron radii from all calculations with PC-PK1 in the neighborhood of $^{22}$Al are plotted in the last panel against the respective proton and neutron Fermi energies. 
Again a smooth trend on average is observed towards $^{22-23}$Al, while the deviation is clear for $^{21}$Al. 
The trend along the isotopic chain and leading up to $^{22}$Al is even smoother when using PC-F1 and triaxiality is considered. 

We conclude that the present calculations, which consistently take into account pairing, deformation, and continuum effects, do not support the development of a proton halo in the $^{22}$Al ground state.


\section{Conclusions\label{sec:summary}} 
We have analyzed theoretical results for the ground state of $^{22}$Al obtained within relativistic density functional theory, including self-consistently the effects of pairing, deformation and the continuum, and using different functionals, in order to assess whether a proton halo can develop in this proton-drip line nucleus. 
Although the valence proton of the nucleus is found very loosely bound, its spatial distribution hardly penetrates the potential barrier. 
Its wave function is indeed found to consist predominately of $\ell =2$ components, for which halo formation is disfavored. 
Comparisons with results for isobars reveal a somewhat more extended density distribution than that of the stable or neutron-rich counterparts, but comparisons along isotopic, isotonic, and isobaric chains reveal no discontinuities in size evolution, which, if present, might signal exotic structures. 
We conclude that $^{22}$Al is very unlikely to develop a proton halo in its ground state.

\subsection*{Acknowledgments}
We thank the members of the DRHBc Mass Table Collaboration for useful discussions and comments. 
P.P was supported by the
 Institute for Basic Science (IBS) through the NRF (2013M7A1A1075764); 
M.-H.M. was supported by the National Research Foundation of Korea NRF grants funded by the Korean government Ministry of Science and ICT (Grant Nos. NRF-2021R1F1A1060066); 
K.Z. was supported by the National Natural Science Foundation of China (Grant No. 12305125), the Sichuan Science and Technology Program (Grant No. 2024NSFSC1356), and the National Key Laboratory of Neutron Science and Technology (Grant No. NST202401016). 
\vspace{3mm} 

\noindent
$^1$Shared first authorship: These authors contributed equally to this work. 

%

\end{document}